\pdfoutput=1
\documentclass[12pt]{iopart}
\usepackage[dvipsnames]{xcolor}
\usepackage[normalem]{ulem}
\usepackage{bm}
\usepackage[utf8]{inputenc}
\usepackage{pgfplots}
\usepackage{tikz}
\usepgfplotslibrary{groupplots}
\pgfplotsset{compat=1.14}
\usepackage{capt-of}
\usepackage{graphicx}
\usepackage{amssymb}
\usepackage{amsthm}
\usepackage{mathrsfs}
\usepackage{hyperref}
\usepackage{lineno}
\hypersetup{colorlinks=true, citecolor=blue}

\begin{document}

\title[AG Sullivan et al.]{Can we use Next-Generation Gravitational Wave Detectors for Terrestrial Precision Measurements of Shapiro Delay?}

\author[Sullivan et al./]{Andrew G. Sullivan$^{1}$, 
Do\u{g}a Veske$^{1}$, 
Zsuzsa M\'arka$^{2}$, 
Imre Bartos$^{3}$, 
Stefan Ballmer$^{4}$, 
Peter Shawhan$^{5}$, 
and Szabolcs M\'arka$^{1}$
}
\address{$^{1}$Department of Physics, Columbia University in the City of New York, New York, NY 10027, USA\\
$^{2}$Columbia Astrophysics Laboratory, Columbia University in the City of New York, New York, NY 10027, USA\\
$^{3}$Department of Physics, University of Florida, Gainesville, FL 32611-8440, USA\\
$^{4}$Department of Physics, Syracuse University, Syracuse, NY 13210, USA\\
$^{5}$Department of Physics, University of Maryland, College Park, MD 20742, USA}

\ead{ags2198@columbia.edu}

\begin{abstract}
Shapiro time delay is one of the fundamental tests of general relativity and post-Newtonian theories of gravity. Consequently, its measurements can be used to probe the parameter $\gamma$ which is related to spacetime curvature produced by a unit mass in the post-Newtonian formalism of gravity. To date all measurements of time delay have been conducted on astronomical scales. It was asserted in 2010 that gravitational wave detectors on Earth could be used to measure Shapiro delay on a terrestrial scale via massive rotating systems. Building on that work, we consider how measurements of Shapiro delay can be made using next-generation gravitational wave detectors. We perform an analysis for measuring Shapiro delay with the next-generation gravitational wave detectors Cosmic Explorer and Einstein Telescope to determine how precisely the effect can be measured. Using a rotating mass unit design, we find that Cosmic Explorer and Einstein Telescope can measure the Shapiro delay signal with amplitude signal to noise ratios (SNRs) upwards of $\sim28 $ and $\sim43$ in 1 year of integration time, respectively. By measuring Shapiro delay  with this technique, next-generation interferometers will allow for terrestrial measurements of $\gamma$ in the paramaterized post-Newtonian formalism of gravity with sub-percent precision.
\end{abstract}

%
%
%
%
%

\section{Introduction}
\label{sec:Introduction} 

The general theory of relativity suggests a number of observable properties as a consequence of the equivalence principle and spacetime curvature \cite{will_1993}. Improved tests for general relativity continue to be sought in physics to increase measurement precision and repertoire for testing the theory \cite{2009SSRv..148....3W, 2014LRR....17....4W}. 

One of the long-known tests of these properties lies in measuring the time delay of light as it passes close to massive objects, as first described by Irwin Shapiro in 1964 \cite{1964PhRvL..13..789S}. Shapiro proposed using a radar pulse directed at a planet near the Sun and measuring the delay in the travel time to and from the planet as a result of the presence of the Sun. In the limiting case where the Earth and the planet are on nearly opposite sides of the sun, the one-way time delay for this experiment as predicted by general relativity is given by
\begin{equation}
    \delta t={2GM_\odot\over c^3} \ln{({4r_1r_2\over d^2})} 
\end{equation}
where $G$ is Newton's gravitational constant, $M_\odot$ is the mass of the Sun, $c$ is the speed of light, $r_1$ and $r_2$ are the orbital radii of the Earth and the target planet, respectively, and $d$ is the closest distance the light pulse approaches to the center of the sun, {neglecting higher order terms since $\frac{GM_\odot r_{1,2}}{c^2d^2}\ll 1 $} \cite{will_1993, 1964PhRvL..13..789S}. In Shapiro's original formulation there are a number of practical uncertainties in obtaining an exact answer due to initial conditions such as the planetary speeds and locations \cite{1966PhRv..141.1219S}. For planets in the solar system the time delay is substantial, on the order of $10^{-4}$ seconds \cite{1964PhRvL..13..789S}.

In the weak field limit, alternative gravity theories can be characterized by the parameterized post-Newtonian (PPN) formalism as deviations from Newtonian gravity \cite{will_1993}. The \(\gamma\) parameter in the PPN formalism, which is related to spacetime curvature produced by a unit mass, can be determined through Shapiro delay measurements. 
The generalized formulation for Shapiro time delay in the PPN formalism for any metric theory is given by
\begin{equation}
    \delta t=(1+\gamma){GM\over c^3}\ln{({4r_1r_2\over d^2})}
\end{equation}
In general relativity, specifically, $\gamma=1$. 
A number of experiments and observations have placed limits on the value of $\gamma$ including observations of planetary echos \cite{10.2307/2878595}, the Viking relativity experiment \cite{1979ApJ...234L.219R}, measurements of Shapiro delay from quasars due to the presence of Jupiter \cite{2003PhLA..312..147K, 2003ApJ...598..704F}, lunar laser ranging experiments \cite{2017APS..APR.B3001B}, and observations of Mercury's perihelion orbit \cite{2006cosp...36..182I}. The Cassini spacecraft has made the most precise measurement of $\gamma$ to date, with $(1-\gamma)=(2.1 \pm 2.3)\times 10^{-5}$, which is in agreement with the predictions of general relativity \cite{2003Natur.425..374B}. 

All measurements of Shapiro time delay and $\gamma$ to date have been made on astronomical scales. In 2010, Ballmer et al.\ proposed a method for measuring Shapiro time delay on a terrestrial scale using an Advanced LIGO (aLIGO) \cite{2015CQGra..32g4001L} gravitational wave (GW) detector, which could be accomplished by installing a rotating mass unit (RMU) to modulate the gravitational potential along one arm \cite{2010CQGra..27r5018B}. By building up a periodic change in the curvature of spacetime, delays on the order of $10^{-32}$ seconds produced by an RMU with $10^4$ kg masses were shown to be measurable with an amplitude SNR of \(\sim 8.7\)\ over an integration time of 1 year, where the amplitude SNR is defined as the square root of the ratio of the average signal power to the average noise power. 

In 2015, aLIGO began its first observing run  \cite{2010CQGra..27h4006H, 2015CQGra..32a5014M, 2016PhRvD..93k2004M} and currently, several future interferometric GW detectors are proposed for 2030-2050  \cite{Hild_2011, 2014JPhCS.484a2008V, Hild_2009, 2010CQGra..27s4002P, 2020JCAP...03..050M, 2017CQGra..34d4001A, 2019BAAS...51g..35R, 2017arXiv170200786A}. Proposed detectors in the design stage include Cosmic Explorer (CE) \cite{2017CQGra..34d4001A, 2019BAAS...51g..35R}, the Einstein Telescope (ET) \cite{2010CQGra..27s4002P, 2020JCAP...03..050M}, and space based detector LISA \cite{2017arXiv170200786A}. New detectors with substantially better sensitivity will improve the prospect of measuring GW signals \cite{2017PhRvD..96h4039C} and gravitational effects with greater precision. For example, it has been posited that a unique measurement of Shapiro time delay from an asteroid flyby could be made with LISA \cite{2007CQGra..24.3005C}.

In this paper, we revisit the previous work on measuring Shapiro delay with RMUs and laser interferometers \cite{2010CQGra..27r5018B, Matone_2007, 2011PhRvD..84h2002R} and investigate the improvements enabled by GW detectors which are now being planned. In Section \ref{sec:CEGWDET}, we introduce next-generation interferometric GW detection and the suggested RMU extension. In Section \ref{sec:SignalAnalysis}, we detail the expected time delay signal from the RMU and compare the signal to the sensitivities of CE and ET. In Section \ref{sec:discussion}, we discuss our results and the precision of the measurements we expect to make with this experiment. Finally in Section \ref{sec:conclusion}, we consider the implications of  Shapiro delay measurements with next-generation GW detectors and conclude.

\section{Next-Generation Detectors and the Rotating Mass Unit Setup}
\label{sec:CEGWDET}

\begin{figure}[h]
    \centering
    
    \includegraphics[width=11cm]{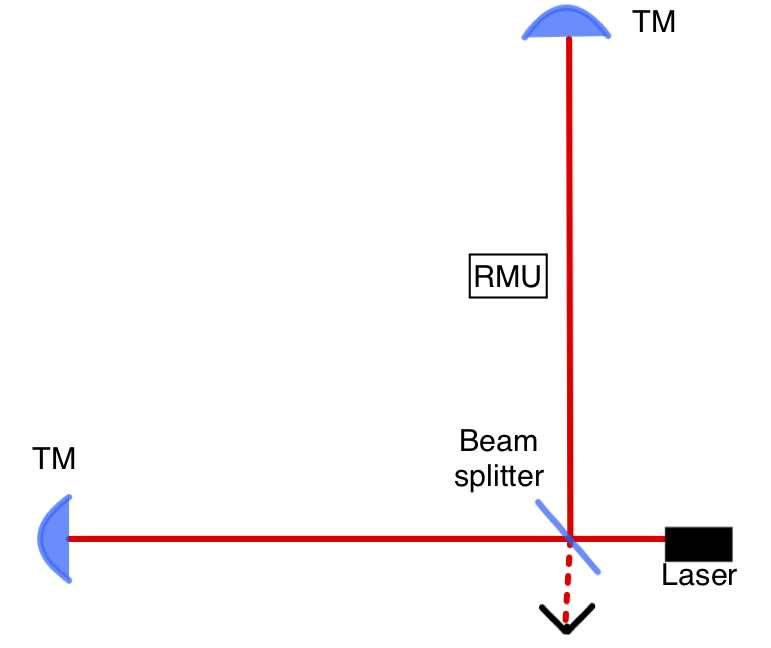}
    \caption{A diagram of a simplified Michelson interferometer design with test masses (TM) for GW detection, including the location of the proposed RMU in the center of one of the interferometer arms.}
    \label{fig:interferdesign}
\end{figure}

Long-arm interferometers operate by making measurements of differences in the length of the interferometer laser arms, making them ideal for measuring GWs, which change the spatial distances during propagation in the weak field limit. GW observation functions via measuring the differential change in the lengths of the arms of a Michelson-type interferometer as a result of a passing GW \cite{2011LRR....14....5P}. Laser interferometers serve to achieve sensitivities capable of detecting GWs on Earth with dimensionless strains on the order of $10^{-21}$ and smaller \cite{2011LRR....14....5P}. Interferometers are sensitive to time-varying signals. Therefore, in order for the Shapiro time delay to be measured, the delay must be modulated \cite{2010CQGra..27r5018B}. A RMU provides a 
way to modulate the Shapiro delay because the rotation allows for a periodic change in the mass density that produces a phase-coherent signal in the interferometer data stream. Figure \ref{fig:interferdesign} depicts the basic design concept of a laser interferometer for GW detection, including the location of the proposed RMU.

CE and ET are the proposed next-generation successors to current generation interferometers aLIGO and Advanced Virgo (AdVirgo) \cite{2015CQGra..32b4001A} in the search for GWs, with much improved sensitivity over aLIGO and AdVirgo \cite{2020JCAP...03..050M, 2017CQGra..34d4001A, 2019BAAS...51g..35R}. CE utilizes a similar Michelson interferometer design shown in Figure \ref{fig:interferdesign}, including Fabry-Perot cavities and signal recycling mechanisms that increase sensitivity. In contrast, ET's most current design ET-D is a triangular shaped composite detector, made up a high frequency (ET-D-HF) and a low frequency (ET-D-LF) detector which are located parallel to one another in the triangle. The two component detectors are further divided into three Michelson interferometers whose arms form a 60 degree angle rather than the 90 degree angle shown in Figure \ref{fig:interferdesign} \cite{Hild_2011}. These parallel detectors will have different specifications in order to achieve their sensitivity goals, with the low frequency detector utilizing a low power laser to reduce acceleration noise and the high frequency detector utilizing a high power laser to reduce quantum shot noise. The goal of these next-generation interferometers is to have strain sensitivity noise floors a factor of 10 greater than the current aLIGO and AdVirgo detectors \cite{2020JCAP...03..050M, 2019BAAS...51g..35R, 2010CQGra..27h4007P} as well as expanded sensitivity at lower frequencies, with CE being sensitive at frequencies as low as 10 Hz and ET being sensitive even lower \cite{Hild_2011, 2020JCAP...03..050M}. CE will have arm cavities that are $40$ km in length, a factor of 10 greater than the length of aLIGO's arm cavities at $4$ km, while each of ET's component interferometers will have 10 km long arm cavities. In addition to cavity length, CE is expected to employ a number of techniques to reduce noise, including increased mirror reflectivity, increased laser power, and increased test mass \cite{2017CQGra..34d4001A, 2019BAAS...51g..35R}. At the low frequency limits, CE {and ET} will likely be limited by Newtonian gravity gradient and seismic noise \cite{Hild_2011, 2017CQGra..34d4001A}, while
exhibiting greater sensitivity in that range than aLIGO \cite{2017CQGra..34d4001A}. 

\begin{figure}[h]
    \centering
    \includegraphics[width=11cm]{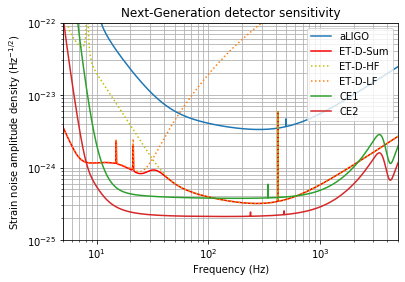}
    \caption{The proposed one-sided strain noise amplitude density curves for next-generation GW detectors. This plot includes both stages of the CE GW observatory \cite{2017CQGra..34d4001A, 2019BAAS...51g..35R}, {the total curve for ET-D, labeled ET-D-Sum as well as its component detectors ET-D-HF and ET-D-LF \cite{Hild_2011},} and aLIGO \cite{2015CQGra..32g4001L} at design sensitivity. The sensitivities shown for ET assumes a 90 degree detector with ET's specifications.}
    \label{fig:sensitivity curve}
\end{figure}

Figure \ref{fig:sensitivity curve} shows the sensitivity curves for CE and ET, including the sensitivity curve of aLIGO for reference. CE will develop in two stages: CE1 (2030s) and CE2 (2040s), where CE2 will be its design sensitivity \cite{2019BAAS...51g..35R}. For this study we adopt the sensitivity curves of CE2 and ET-D-HF, as the improved sensitivity at low frequencies even in ET-D-HF will allow us to employ an RMU that rotates at lower frequencies and will provide a lower noise floor at each signal's primary harmonic than was the case for aLIGO \cite{2010CQGra..27r5018B}.

The proposed RMU would operate as follows: a bar structure suspended by its middle with masses concentrated at the ends of the bar would be placed alongside the center of one of the arms {of these two detectors} and rotate in the plane perpendicular to the laser beam, as described in Ballmer et al.\  \cite{2010CQGra..27r5018B}. As the assembly rotates, the curvature of spacetime through which the laser beam passes changes due to variation in the distance to the masses, thus producing periodic Shapiro delay. To improve precision, a line of synchronized parallel RMUs arranged alongside and orthogonal to the center of one of the GW detector arms may be used (See Figure 1 in Ballmer et al.\ \cite{2010CQGra..27r5018B}). The RMU's rotation would be constantly monitored, ensuring that the data analysis of the Shapiro delay can be done with excellent phase-coherence and precision over long periods of time. 

\begin{figure}[h]
    \centering
    \includegraphics[width=11cm]{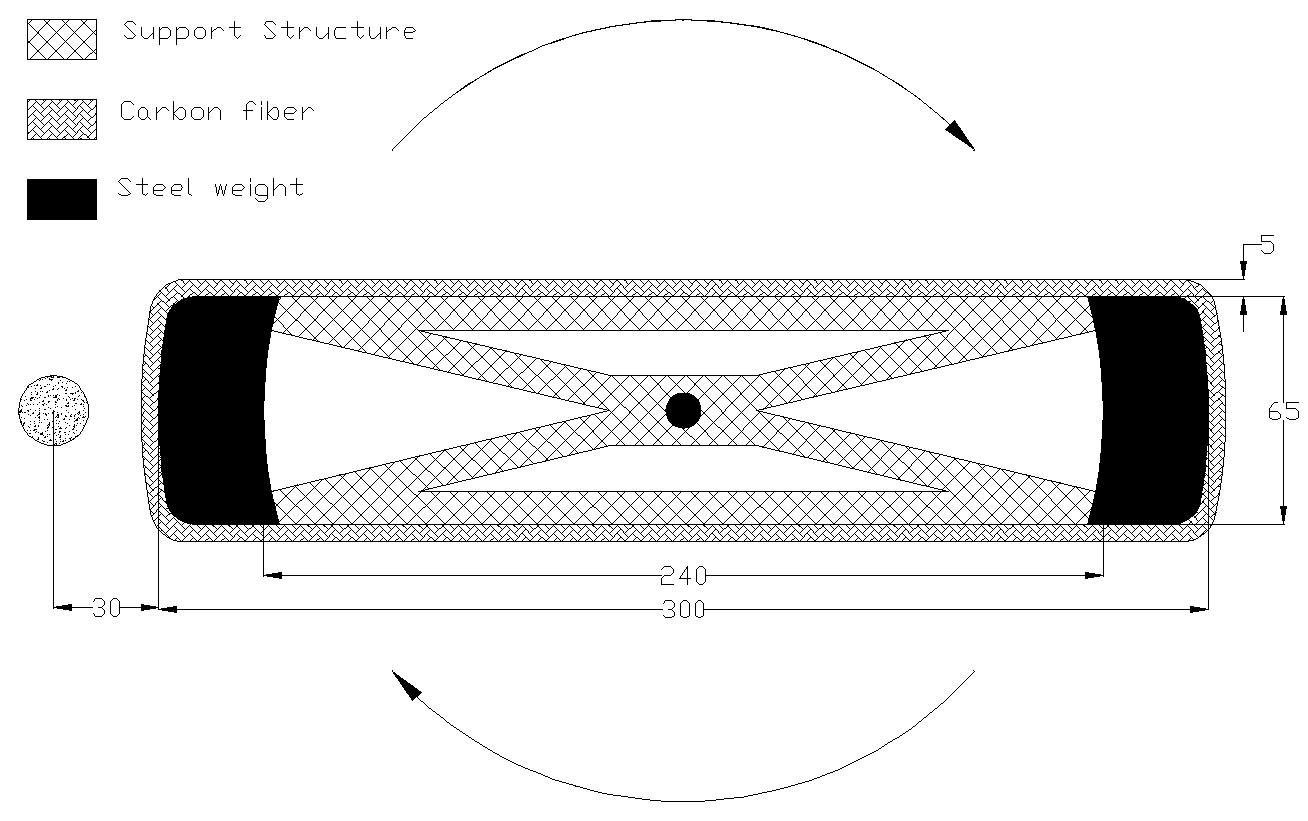}
    \caption{A sketch of the RMU proposed by Ballmer et al.\ 
    \cite{2010CQGra..27r5018B}. Each dimension is measured in centimeters. The circle to the left of the rotating mass represents the interferometer laser beam, placed 30 centimeters from the closest point of the mass. We adopt the same symmetric design, but vary the steel mass size to either reduce the expense or maximize the scientific output of such a device. Figure taken from Ballmer et al.\ Figure 3 \cite{2010CQGra..27r5018B}.} 
    \label{fig:rotatingmass}
\end{figure}

We consider whether a symmetric RMU (i.e. one where the center of mass is located at the center of the bar) is the optimal geometry for measuring the Shapiro delay. The time delay is proportional to a line integral of the Newtonian potential $U$. The Newtonian potential behaves $\propto {M\over{r}}$, where $M$ is the mass generating the gravitational field and $r$ is the distance from the mass to the point of interest. For an asymmetric rotor system, the ratio of rotation radii of the end masses would need to be proportional to the inverse ratio of the end masses to maintain a stable axis of rotation, thus making the smaller mass's minimum distance to the interferometer laser beam closer than the larger mass's minimum distance. Keeping constant the minimum distance to the beam, total mass, and total rotor length, the maximal contribution to the gravitational potential from the smaller mass in the asymmetric rotor reduces by the factor of the mass ratio when compared with the symmetric rotor since its minimum distance to the beam doesn't change. However the maximal contribution from the large mass doesn't increase by the same factor since its minimum distance to the beam will be larger. Hence we expect the symmetric design to produce the highest SNR and adopt the symmetric RMU geometry which was used in Ballmer et al.\  \cite{2010CQGra..27r5018B}. Figure \ref{fig:rotatingmass} shows a sketch of the symmetric RMU proposed in Ballmer et al.

For our analysis, we propose three possible RMU models using the symmetric geometry described above: an advanced science model, the Ballmer et al.\ model, and a cost conscious model. We label these three models A, B, and C for convenience. All three models are $3.0$ meters in length \cite{2010CQGra..27r5018B}. The distance of closest approach to the laser beam remains at $30$ cm  as depicted by Figure \ref{fig:rotatingmass}. We design model A to have end masses of $2.5\times 10^4$ kg and rotate at a frequency of $25$ Hz to most effectively induce detection as it places the fundamental and subsequent signal frequency components in the most sensitive frequency range of both CE and ET.  To account for the greater steel mass in model A, we propose extending the width of the rotor in the direction parallel to the laser beam to satisfy size constraints. Model B is the exact one described in the example presented by Ballmer et al.\ with $1.5 \times 10^{4}$ kg steel weight end masses, rotating at a frequency of $25$ Hz. Model C possesses end masses of $6500$ kg and rotates at a lower frequency of $15$ Hz, requiring substantially less energy to be stored in the heavy rotating masses, and therefore decreasing material strength and cost requirements. A lower rotational frequency and mass better ensure the safety of operation. Additionally, the substantially improved sensitivity of both CE and ET over aLIGO at lower frequencies allows for a signal to be measured whose primary harmonic is below 50 Hz.

To construct these RMUs a number of constraints exist, particularly in rotating such heavy steel masses at frequencies exceeding $15$ Hz. Model A will require 690 MJ of energy stored in each half of the RMU to rotate the system, while models B and C will require 420 MJ and 65 MJ each. It should in principle be possible to accomplish this feat by constructing the RMU with a light weight structure and a carbon fiber frame and decomposing the RMU into sections supported by multiple bearings and operated by multiple motors as considered in Ballmer et al.\ \cite{2010CQGra..27r5018B}. Prior to construction, a finite element analysis would need to be conducted on the RMU to quantitatively study the stresses on the system over time. More specific studies into the construction of this system are beyond the scope of this paper. 

Additionally, a number of factors will constrain use of the RMU once constructed. Narrowband features at various frequencies in the amplitude spectral density of the real detector, caused by power lines, mechanical resonances, and calibration noise \cite{PhysRevD.97.082002, PhysRevLett.116.061102, LIGOsensDCC}, must be accounted for in selecting the frequency of the RMU. To avoid these lines interfering with the Shapiro delay signal, one can simply modify the frequency of the RMU so that it and its subsequent harmonics do not overlap with the lines.  Additionally, it may be possible for the RMU to interfere with the detection of continuous astrophysical GW sources which may be discovered in future observing runs \cite{PhysRevD.100.024004, PhysRevLett.124.191102, Papa_2020}. If this should occur, the RMU frequency can either be changed or the RMU can simply be turned off so that the potential continuous GW signal may be distinguished. For this study, we do not consider the interferometer lines beyond those present in the design sensitivity curves for ET and CE.

One additional consideration we account for in calculating the Shapiro delay is the gravitational coupling attraction between the RMU and the TM at the end of the laser arm. Ballmer et al.\ make this consideration, but do not account for the resonances of the TM suspension. Taking $L$ to be the length of the detector arm, this acceleration behaves $\propto L^{-6}$ (See the Appendix for details). Solving for the position of the interferometer TM as a function of time assuming the TM to be a harmonic oscillator with resonance frequency $f_r$, we obtain the amplitude of oscillation for each of the three spatial degrees of freedom. The amplitude of oscillation in the axis along the laser pathway d$z$ is as follows:
\begin{equation}
\textrm{d}z = \frac{384}{4\pi^2}{GMl^{2}(d+l)^{2}\over{L^{6}(f_{r}^{2}-4f^2)}}
\label{eq:zamplitude}
\end{equation}
where $M$ is the RMU end mass, $d$ is the closest distance the {center of the} RMU end mass approaches to the interferometer laser beam, $l$ is half the length of the RMU, and $f$ is the frequency of the RMU. The frequencies chosen for all 3 models exceed the greatest resonance frequency of the TM suspensions for aLIGO \cite{LIGOsusom}, making the $(f_{r}^{2}-4f^2)^{-1}$ factor no greater than order $\sim1$ s$^2$, since the resonance frequencies for next-generation detectors will be even lower. Thus, the $\textrm{d}z$ coupling can be neglected, as L is {of order $10^4$} m for next-generation detectors, making the effect on the order of at most 100 Planck lengths. In the directions perpendicular to the laser path (i.e. $\textrm{d}x$ and $\textrm{d}y$), the coupling behaves $\propto L^{-5}$ (see Appendix); however, this coupling has a negligible effect on the motion of the TM in the direction of the laser arm  \cite{freise2010interferometer}. 
\section{Signal Analysis}
\label{sec:SignalAnalysis}
 The time delay of a light signal passing near a massive object is given by
\begin{equation}
\delta t = \int (1+\gamma)U \textrm{d}s
\label{eq:timedelay}
\end{equation}
in the weak field limit \cite{will_1993} where the integral is over the signal path and \(U\) is the Newtonian potential scaled with the reciprocal of the cube of speed of light as follows:
\begin{equation}
     U = \sum_{i=1}^{N}{GM_i\over{c^3r_i}}
\end{equation}
for $N$ masses, where $c$ is the speed of light and $r_i$ is the distance to each mass $M_i$. For this arrangement of one RMU, it corresponds to the distance from the interferometer's laser beam to the masses at the ends of the rotor, and $N=2$. To produce a signal with a larger amplitude one can determine the potential for an arrangement of multiple RMUs. For this principle study we calculate the signals produced by one RMU.

To evaluate the time delay for this proof of concept demonstration, we simplify the geometry of each RMU for our calculations, assuming the end masses to be {spherically symmetric with radii of 20 cm} and the support systems holding them up to have negligible mass as the end masses will be substantially larger than the support system. We note that this will likely yield a conservative estimate as much of the mass will be further away from the beam than is shown in Figure \ref{fig:rotatingmass} (Ballmer et al.\ do not make this assumption in their paper, leading to a difference in SNR of about 20\% from what we report here). 

By evaluating the integral in Equation \ref{eq:timedelay} {over the arm-lengths of both CE and ET} for each of our three models, we obtain the Shapiro time delay as a function of time. The Shapiro time delay from a general RMU configuration in one laser arm is
\begin{equation}
  \delta t=2(1+\gamma)\frac{GM}{c^3}{\textrm{arcsinh}{L^2 \over 4d^2 \sqrt{(1+{2l\over d}+\frac{2l^2}{d^2})^2-{4l^2(d+l)^2\over d^4}\cos^2{2 \pi f t}}}} \label{eq:shap}
\end{equation}
where $L$ is the interferometer arm-length, $l$ is half the length of the RMU, $d$ is the closest distance the RMU approaches to the laser beam plus the radius of the end mass, $M$ is the end mass of the RMU, and $f$ is the frequency of oscillation of the RMU. 
We require all RMU models to have $2l=3.0$ m and $d=50$ cm, 30 cm from the minimum distance to the beam and $20$ cm from the radius of the spherical end mass. 

Because ET will be composed of multiple different interferometers working in coincidence, there are a number of possible laser beams to situate the RMU alongside. In practice, one must consider whether to place the RMU alongside the ET-D-HF beam, the ET-D-LF beam, or the space between the two beams to produce the same signal in both detectors. Additionally, it may be possible to increase the Shapiro delay signal's total SNR by a factor of $\sqrt{2}$ by rotating the RMU inside the arm cavity between the two parallel interferometer laser beams; however, due to the uncertainty in the exact dimensions of the arm cavities at the time of this paper's writing and the fact that the RMU frequencies chosen will produce signals predominantly in the high frequency band, we calculate the signal in only one ET-D-HF laser beam nearest to the RMU, choosing the beam to be 50 cm away from the center of the spherical end mass. 

\begin{figure}[h]
    \centering
    \includegraphics[width=11cm]{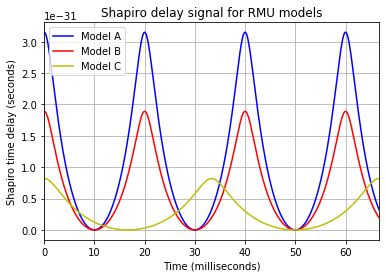}
    \caption{The oscillating component of the time delay in CE as a function of time for all models over $67$ milliseconds, the period of rotation for model C. Models A and B rotate at 25 Hz and model C rotates at 15 Hz.}
    \label{fig:timedelay}
\end{figure}

Time delays for our three models are shown in Figure \ref{fig:timedelay}. {Only the time delay for CE is shown in Figure \ref{fig:timedelay} as the difference in arm-length between CE and ET has a negligible impact on the time varying component of the Shapiro delay.} The time delay exhibits variations on the order of $10^{-31}$ seconds for models A and B, and $10^{-32}$ seconds for model C. Multipying by a factor of the speed of light, this corresponds to an optical distance change on the order of $10^{-23}$ meters for models A and B and $10^{-24}$ for model C, displacements that both interferometers will have the capability to measure. 

Because of the triangular shape of ET and its component interferometers, using the noise curve to calculate SNR is in general nontrivial  \cite{Hild_2011}. For the Shapiro delay case, however, the time delay produced by the RMU in the far laser arm will be negligible since the amplitude of oscillation falls off strongly with distance to the beam as Equation \ref{eq:shap} indicates. Consequently, we can treat ET as a detector whose arms are perpendicular for this analysis since only one of the laser arms will be affected by the Shapiro delay.

We calculate the SNRs over 1 year for the three models with the proposed noise curves for CE and ET. Since we only calculate the ET Shapiro delay in the ET-D-HF beam, we calculate the SNRs using ET-D-HF sensitivity rather than accounting for the effects of the other interferometers working in coincidence. As a result of this approximate treatment our ET SNRs are likely conservative estimates. 

To obtain the signal that would appear in the interferometer datastream, the time delay is converted to an equivalent differential light path displacement. Since dimensionless strain $h(t)$ is defined as the differential change in the length of the light path divided by the resting interferometer arm-length, the strain produced by the Shapiro delay is
\begin{equation}
    h(t)={c\delta t(t) \over L }
\end{equation}
where $c$ is the speed of light, $\delta t(t)$ is the time dependent component of the Shapiro delay, and $L$ is the length of the interferometer arm. {As the quantity of interest is time delay, we use the dimensionless strain purely as a calculating tool to determine the SNR.} We calculate the SNR of this strain signal by using matched filtering and assuming our signals are periodic, so that their Fourier transforms have impulses. {We compute the individual Fourier components numerically and calculate their individual SNRs. The total SNR for CE is obtained from the individual SNRs of the first five harmonics as this exceeds the number of harmonics that have SNRs greater than 1 for all RMU models with CE, while the total SNR for ET is obtained from the first 8 harmonics.} We quote 1 year of integration time results as a basis since SNR grows proportionally to the square root of time. (See \footnote{The SNR grows with the square root of time. As such for longer observation times, we expect an even more substantial increase in the precision of this measurement. Our model A will generate an SNR of 64 in 5 years and 90 in 10 years {with CE}. These SNRs correspond to standard errors in the measurement of $\gamma$ of $\pm3.1\%$ and $\pm2.2\%$. {With ET, model A will generate an SNR of 97 in 5 years and 137 in 10 years, yielding $\gamma$ measurements with standard errors of $\pm2.1\%$ and $\pm1.5\%$}} for note on long data taking). 

\begin{center}
\begin{table}[h!]
\setlength{\tabcolsep}{16pt}
\begin{tabular}{ p{6cm} c c c  }
 
 \hline\hline
 Model     & \textbf{A} & \textbf{B} & \textbf{C}\\
 \hline\hline
 End Mass (kg) & $2.5\times10^{4}$ & $1.5\times10^4$ & $6.5\times10^3$ \\
 Rotational Frequency (Hz) & 25 & 25 & 15 \\
 \hline
 CE Total SNR & 29 &17& 7.2 \\
 CE Primary Harmonic SNR   &27 & 16 &6.8\\
CE Secondary Harmonic SNR & 7.8 & 4.7 & 2.0\\
\hline
  ET Total SNR & 43 &26& 7.0 \\
 ET Primary Harmonic SNR   &39 & 24 &5.9\\
ET Secondary Harmonic SNR & 17 & 10 & 3.4\\
 \hline\hline
\end{tabular}
\caption{The 1 year total SNR and SNRs for the primary and secondary harmonics in both CE and ET for each of the three models. We include the end mass and rotational frequency of each RMU model for referential convenience.}
\label{table:1}
\end{table}
\end{center}

Our results are listed in Table \ref{table:1}. In the case of CE, we find that the total SNR for 1 year of integration time to be 29 for model A. Additionally, model B yields an SNR of 17, and model C yields an SNR of 7.2. For model A, 
the fundamental frequency at $50$ Hz (the factor of 2 comes from the mass symmetry) is observable with an SNR of $27$, while the second and third harmonics have SNRs of $7.8$ and $2.9$ (including both the positive and negative frequency components), respectively, over 1 year of integration time. 
For the signal produced by model B, the SNR for the fundamental frequency is $16$; the second and third harmonics have SNRs of $4.7$ and $1.7$ respectively for 1 year of integration time. Model C generates a signal whose fundamental frequency will be observable with an SNR of $6.8$, and second harmonic with an SNR of $2.0$ in 1 year of integration time. We find the SNRs expected from making this measurement with ET in 1 year to be more substantial. The total SNR calculated from model A with ET is 43, while models B and C yield SNRs of 26 and 7.0. More harmonics are detectable with ET than CE as the first 5 harmonics are detectable of the model A signal with ET. The first 4 harmonics from model B are detectable while the first 3 harmonics are detectable from model C in 1 year.

\begin{figure}[h]
   \centering
    \includegraphics[width=11cm]{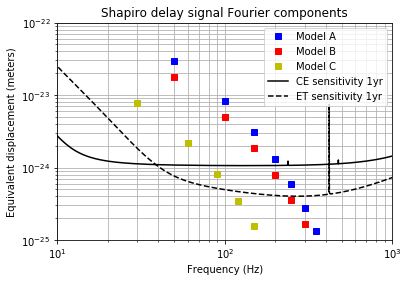}
    \caption{The Fourier series coefficients of the Shapiro delay signal converted to the equivalent light path displacement as a result of the delay produced by each of the three RMU models. Also included are the two-sided noise amplitude curves for CE2 and ET-D-HF over 1 year of integration time, scaled to be in units of displacement.
    }
    \label{fig:fouriercomp}
\end{figure}

We show the Fourier series coefficients of each model's signal and compare them to the CE and ET sensitivity curves for $1$ year of integration time in Figure \ref{fig:fouriercomp}. {As strain is used simply for SNR calculation}, we present the signal components in terms of the equivalent light path displacement and multiply the noise curves shown in Figure \ref{fig:sensitivity curve} by a factor of $L/\sqrt{2T}$ where the factor of $\frac{1}{\sqrt{2}}$ converts the one-sided amplitude spectral density to a two-sided amplitude spectral density and $T$ is integration time to obtain the amplitude curve in units of observable displacement. Although ET has a higher noise floor in dimensionless strain than CE as shown in Figure \ref{fig:sensitivity curve}, when scaled with a factor of arm-length its floor is lower than CE's because ET's arms are a factor of 4 shorter than CE's. Note that model C has Fourier series coefficients at different frequencies than the other two models because it oscillates at 15 Hz rather than 25 Hz, making the Fourier components multiples of 30 Hz rather than 50 Hz. 

\section{Discussion}
\label{sec:discussion}

The prior analysis showed that aLIGO has has the capability to probe Shapiro delay on a terrestrial level; however, a rotor with $1.5\times 10^4$ kg end masses would only yield an SNR of 6.5 in 1 year of integration time (making {the 20 cm radius spherical} mass assumption). Observing the Shapiro delay with that SNR would allow for a measurement of $\gamma$ with  $\sim \pm25\%$ precision for one standard deviation \cite{2010CQGra..27r5018B}. 

Results for our three RMU models demonstrate the new experimental gravity test that can be feasibly achieved with next-generation ground-based detectors. Our model A with $2.5\times 10^{4}$ kg end masses produces a Shapiro delay with a total SNR over $\sim29$ with CE and over $\sim43$ with ET in 1 year, noticeably better than the scenario considered by Ballmer et al.
By directly comparing the signal results produced by the model B to the results from Ballmer et al., we find that both CE and ET yield superior measurement capabilities to aLIGO, as CE will record a total SNR of $\sim17$ and ET will record a total SNR of $\sim26$ rather than $\sim6.5$ in 1 year of integration time. This is $\sim2.6$ and $\sim4$ times larger. We note that although CE2 and ET are over a factor of 10 more sensitive than aLIGO in strain as shown in Figure \ref{fig:sensitivity curve}, the larger arm-lengths of next-generation detectors with respect to aLIGO reduce the time delay signal's  dimensionless strain amplitude and prevent the SNR obtained with the next-generation detectors from being a factor of 10 greater.

Model C possesses the most conservatively sized masses: less than half of that of the rotor system used in the Ballmer et al.\ calculation, and still produces a signal with an SNR of 7.2 with CE and 7.0 with ET. This would achieve a similar level of precision expected by Ballmer et al.\ in 1 year of integration time, but, because of CE and ET's advanced sensitivities, will cost less material and energy to operate and improve measurement precision as the integration time increases. Additionally, our results for the low cost model demonstrate that the rotor can operate at a lower frequency and still produce appreciable results. Because of CE's flat noise curve from 30 Hz onward as well as ET's lower noise floor and the ability to place an RMU next to the ET-D-LF interferometer, utilizing an RMU that rotates at a lower frequency would be feasible.

ET achieves noticeably larger SNR measurements for models A and B than CE despite having a higher strain noise floor. This is a consequence of strain from the Shapiro delay being a factor of 4 larger in ET than in CE as the arm-length is 4 times shorter. Additionally, the lower metrology noise and dip in noise amplitude at high frequencies in the ET-D-HF detector allows for the detection of the higher frequency harmonics with larger SNRs.

Assuming additive zero mean Gaussian noise, the noise at the output of the matched filter and consequently the total output will be Gaussian. Therefore, the delay measurements are expected to yield a Gaussian distributed measurement of (1+$\gamma$) whose mean is the average filtered signal amplitude and standard deviation is the standard deviation of the filtered noise probability distribution function. The amplitude SNR will be the ratio of the mean of the Gaussian to its standard deviation. We can thus find the standard errors in the estimation of both the Shapiro delay and $\gamma$ directly from SNRs. In real GW detectors the noise is not purely Gaussian, mainly due to the Poisson-like impulsive noise known as glitches  \cite{PhysRevD.71.062001}. In astrophysical {transient} searches the glitches can produce a comparable SNR to the SNR of an astrophysical compact binary coalescence  \cite{Abbott_2016}. Therefore SNR is not used as the only threshold test statistic in those searches  \cite{PhysRevD.71.062001}. However, in our case we do not use SNR as a threshold statistic {to answer the yes/no question of whether there is a real signal in the detector; rather, we use it to estimate a real valued quantity. In our case, the effect of glitches would be the distortion of the Gaussianity of the filtered output and a change in the estimated noise PSDs. Nevertheless, as a known issue it can be mitigated by data cleaning if the glitch models are well understood, or simply by omitting the regions with glitches.} 

With model B, the model used in Ballmer et al., CE can obtain a measurement of the  Shapiro delay with up to $\pm\sim5.8\%$ standard error and $\gamma$ with up to $\pm\sim12\%$ standard error. ET can achieve greater precision, measuring Shapiro delay from model B with a standard error of $\pm3.8\%$ and $\gamma$ with a standard error of $\pm7.7\%$ in 1 year. Using our advanced science model A, we expect to achieve measurements of $\gamma$ with a {standard error of $\pm6.9\%$} with just 1 year of observation with CE and a standard error of $\pm4.6\%$ with ET. Precision also increases with time\footnotemark[\value{footnote}], so over longer times and with the use of multiple synchronized RMUs, this technique can achieve sub-percent precision measurements. With an arrangement of two synchronized model A RMUs, ET can measure $\gamma$ with standard error below $\pm1\%$ in just over 5 years, while three synchronized model A RMUs can allow CE to measure $\gamma$ with standard error under $\pm1\%$ in the same interval.

\section{Conclusion}
\label{sec:conclusion}

We presented an analysis that details the use an RMU with the next-generation GW detectors CE and ET to measure Shapiro time delay. Despite all prior Shapiro time delay measurements being space-based experiments, we show that RMU models are capable of generating a substantial Shapiro delay signals measurable on Earth using these next-generation detectors. 
As detailed in our discussion, our RMU produces a signal with an appreciable SNR in 1 year of observation with both detectors. 
This corresponds to the most precise proposed Earth-based measurement scheme of $\gamma$ to date with a noticeable increase in the measurement precision achievable with aLIGO, as CE and ET can measure $\gamma$ with standard errors below $\pm\sim7\%$ in just 1 year with the installation of one RMU. 

 This technique has its limitations, as it does not yet approach the level of precision achieved by longer range, space-based experiments such as that of the Cassini spacecraft, which achieved a precision of $\sim0.005\%$ in 6 years of observation \cite{2003Natur.425..374B}. In 6 years of observation time with ET and one RMU, this technique would yield a precision of $\pm1.9\%$. An array of synchronized RMUs, however, would increase the precision, achieving sub-percent measurements in half a decade. Nevertheless, next-generation GW detectors promise to be capable of probing exciting gravity science including precise measurements of Shapiro delay and $\gamma$ on Earth.
\section*{Acknowledgments}
The authors thank Columbia University in the City of New York, University of Florida, Syracuse University, and University of Maryland for their generous support.
The Columbia Experimental Gravity group is grateful for the generous support of the National Science Foundation under grant PHY-1708028. AS is grateful for the support of the Columbia College Science Research Fellows program. DV is grateful to the Ph.D. grant of the Fulbright foreign student program and the Jacob Shaham Fellowship. IB acknowledges the support of the Alfred P. Sloan Foundation. PS is grateful for the support of the National Science Foundation through grant PHY-1710286. SB is grateful for the support of the National Science Foundation through grant PHY-1836702.

\appendix

\section{Newtonian Coupling with the Test Mass}
\label{appendix:a}
The Newtonian coupling between the RMU and the TM is obtained by paramaterizing the Newtonian gravitational force in terms of time in three spatial directions. Assuming the RMU is at the center of the laser arm, the three components of the Newtonian force on the TM are
\begin{equation}
    F_z={-4GMm \over L^2}({1 \over ((1+4({d+l\over L})^2+{4l^2\over L^2})+{8l(d+l) \over L^2}\cos{\omega t})^{3/2}}+{1 \over (1+4({d+l\over L})^2+{4l^2\over L^2}-{8l(d+l) \over L^2}\cos{\omega t})^{3/2}})
\end{equation}
\begin{equation}
    F_x={8GMm \over L^3}({d+l(1+\cos{\omega t}) \over (1+4({d+l\over L})^2+{4l^2\over L^2}+{8l(d+l) \over L^2}\cos{\omega t})^{3/2}}+{d+l(1-\cos{\omega t}) \over (1+4({d+l\over L})^2+{4l^2\over L^2}-{8l(d+l) \over L^2}\cos{\omega t})^{3/2}})
\end{equation}
\begin{equation}
    F_y={8GMm \over L^3}({l \sin{\omega t} \over (1+4({d+l\over L})^2+{4l^2\over L^2}+{8l(d+l) \over L^2}\cos{\omega t})^{3/2}}-{l \sin{\omega t} \over (1+4({d+l\over L})^2+{4l^2\over L^2}-{8l(d+l) \over L^2}\cos{\omega t})^{3/2}})
\end{equation}
where $z$ is the axis along the laser beam path, $x$ is the axis formed by the laser beam and the center of the RMU, $y$ is the axis perpendicular to both $x$ and $z$, $M$ is the end mass of the RMU, $m$ is the TM, $l$ is half the length of the RMU, $d$ is the minimum distance between the RMU and the laser beam, $L$ is the length of the detector arm, and $\omega=2\pi f$ where $f$ is the frequency of the RMU. Combining the denominators of all three forces, one obtains
\begin{equation}
    F_z={-4GMm \over L^2}({(1+4({d+l\over L})^2+{4l^2\over L^2}+{8l(d+l) \over L^2}\cos{\omega t})^{3/2} +(1+4({d+l\over L})^2+{4l^2\over L^2}-{8l(d+l) \over L^2}\cos{\omega t})^{3/2} \over ((1+4({d+l\over L})^2+{4l^2\over L^2})^2-({8l(d+l) \over L^2})^2\cos^2{\omega t})^{3/2}})
\end{equation}
\begin{equation}
    F_x={8GMm \over L^3}({(d+l(1+\cos{\omega t}))(1+4({d+l\over L})^2+{4l^2\over L^2}-{8l(d+l) \over L^2}\cos{\omega t})^{3/2})\over ((1+4({d+l\over L})^2+{4l^2\over L^2})^2-({8l(d+l) \over L^2})^2\cos^2{\omega t})^{3/2}} + {(d+l(1-\cos{\omega t}))(1+4({d+l\over L})^2+{4l^2\over L^2}+{8l(d+l) \over L^2}\cos{\omega t})^{3/2} \over ((1+4({d+l\over L})^2+{4l^2\over L^2})^2-({8l(d+l) \over L^2})^2\cos^2{\omega t})^{3/2}})
\end{equation}
\begin{equation}
    F_y={8GMm l \sin{\omega t} \over L^3}({(1+4({d+l\over L})^2+{4l^2\over L^2}-{8l(d+l) \over L^2}\cos{\omega t})^{3/2} -(1+4({d+l\over L})^2+{4l^2\over L^2}+{8l(d+l) \over L^2}\cos{\omega t})^{3/2} \over ((1+4({d+l\over L})^2+{4l^2\over L^2})^2-({8l(d+l) \over L^2})^2\cos^2{\omega t})^{3/2}})
\end{equation}
Assuming that $L>>d,l$, we apply binomial approximations to both the numerator and denominator of all the forces. After combining like terms and neglecting the constant component of the force, we arrive at expressions for the periodic force terms of highest order in $L$ for each of the three translational degrees of freedom:
\begin{equation}
    F_z=\frac{384 GmMl^2(d+l)^2}{L^6}\cos{2\omega t}
\end{equation}
\begin{equation}
    F_x=\frac{96 GmMl^2(d+l)}{L^5}\cos{2\omega t}
\end{equation}
\begin{equation}
    F_y=\frac{96 GmMl^2(d+l)}{L^5}\sin{2\omega t}
\end{equation}

\section*{References}
\bibliography{Refs} 

\end{document}